# Melting and thermal ablation of a silver film induced by femtosecond laser heating: A multiscale modeling approach

Pengfei Ji and Yuwen Zhang[1]

Department of Mechanical and Aerospace Engineering
University of Missouri
Columbia, MO 65211, USA

## Abstract

The femtosecond laser pulse heating of silver film is investigated by performing quantum mechanics (QM), molecular dynamics (MD) and two temperature model (TTM) integrated multiscale simulation. The laser excitation dependent electron thermophysical parameters (electron heat capacity, electron thermal conductivity, and effective electron-phonon coupling factor) are determined from *ab initio* QM calculation, and implemented into TTM description of electron thermal excitation, heat conduction, as well as electron-phonon coupled thermal energy transport. The kinetics of atomic motion is modeled by MD simulation. Energy evolution of excited electron subsystem is described by TTM in continuum. The MD and TTM are coupled by utilizing the effective electron-phonon coupling factor. Laser heating with varying laser fluence is systematically studied to determine the thresholds of the homogeneous melting and ablation. The thermal ablation induced by faster expansion of locally and excessively superheated silver is reported. This paper provides a basis for interpreting the phase change process induced by laser heating, and facilitates the advancement of femtosecond laser pulse processing of material.

## Nomenclature

| | |
|---|---|
| $A$ | material constants describing the electron-electron scattering rate, $s^{-1}K^{-2}$ |
| $B$ | material constants describing the electron-phonon scattering rate, $s^{-1}K^{-1}$ |
| $C_e$ | electron heat capacity, $J/(m^3 K)$ |
| $E$ | energy, $J$ |
| $f$ | Fermi-Dirac distribution function |
| $g$ | electron density of states |
| $G_{e-ph}$ | electron-phonon coupling factor, $W/(m^3 K)$ |
| $J$ | Laser fluence, $J/cm^2$ |
| $k$ | thermal conductivity, $W/(mK)$ |
| $k_B$ | Boltzmann constant, $1.38 \times 10^{-23} J/K$ |
| $m$ | mass, $kg$ |
| $L$ | penetrating depth, $m$ |
| $q$ | heat flux, $W/m^2$ |
| $\mathbf{r}_i$ | position of an atom |

---
[1] Corresponding author. Email: zhangyu@missouri.edu

| | |
|---|---|
| $t$ | time, $s$ |
| $T$ | temperature, $K$ |
| $v$ | velocity, $m/s$ |
| $V_c$ | Volume of unit cell, $m^3$ |

**Greek Letters**

| | |
|---|---|
| $\varepsilon$ | electron energy level, $J$ |
| $\mu$ | chemical potential, $J$ |
| $\lambda\langle\omega^2\rangle$ | second moment of the electron-phonon spectral function, $meV^2$ |
| $\rho$ | density, $kg/m^3$ |
| $\tau_e$ | total electron scattering time |
| $\tau_{xx}$ | thermal stress, $GPa$ |

**Subscripts and Superscripts**

| | |
|---|---|
| $ba$ | ballistic transport |
| $e$ | electron |
| $F$ | Fermi level |
| $l$ | lattice |
| $op$ | optical penetration |
| $p$ | laser pulse |

# 1. Introduction

Femtosecond laser pulse processing of material has been drawn considerable attention and been intensively studied [1–4]. It is widely acknowledged as an effective approach in micromachining and microfabrication [5,6]. Comparing with the conventional long laser pulse in material processing, the femtosecond laser has the great merit of small collateral damage of treated material, which results in high precision and resolution. Understanding comprehensive knowledge of the femtosecond laser pulse interaction with material is an essential step in micromachining and microfabrication. With the rapid advancement of laser technology, the laser pulse parameters, such as the laser fluence, beam size, pulse duration, and laser repetition rate can be controlled.

Plenteous approaches were proposed to model the electron thermal excitation and heat diffusion, and electron-phonon coupled heat transfer occurring in the metallic material upon femtosecond laser pulse irradiation. The thermally excited electrons in the metallic material were described by Fermi-Dirac distribution function $f$ [7]. The thermal excitation of electrons accompanying with shifts of electron density of states $g$ and changes of electron chemical potential $\mu$, were crucial parameters impacting the electron heat capacity $C_e$ and effective electron-phonon coupling factor $G_{e-ph}$ [8–12]. The electron thermal conductivity $k_e$ in the laser heated material was derived from the Drude model [13,14]. The two-temperature model (TTM) was originally proposed by Anisimov *et al.* [15] to describe the laser energy absorption by the electron subsystem and thermal energy transferring from the electron subsystem to the lattice subsystem. On the basis of TTM, the Boltzmann transport equation was used by Qiu and Tien in studying the thermal energy transport of electrons and electron-phonon interactions [16]. The lattice Boltzmann method (LBM) was implemented to solve the Boltzmann transport equation in laser heating of silicon film by Mao and Xu [17]. The dual phase lag (DPL) model was proposed by Tzou to describe the femtosecond laser pulse induced phase lag of heat flux in time and phase lag of temperature gradient in space [18]. An interface tracking method proposed by Zhang and Chen [4], which iteratively solves the laser induced material removal on the basis of energy balance and nucleation kinetics, was implemented by Huang *et al.* to study the ultrafast solid-liquid-vapor phase change [19].

Molecular dynamics (MD) simulation provides detailed information on the dynamics evolution and structural change of laser irradiated material [20,21]. Combined atomic and continuum studies of the femtosecond laser pulse processing of material were carried out by describing the electron subsystem via TTM and the lattice subsystem via classical MD. In the combined MD-TTM studies, the electron-phonon coupling was modeled either by imposing an additional energy term in the equation of motion of atoms [22–24], or by scaling up the atomic velocity to represent the thermal energy transferred from the electron subsystem to the lattice subsystem [25]. The combined MD-TTM simulation covered melting [22,25], spallation [23] and ablation [24] upon femtosecond laser pulse irradiation. Pure *ab initio* quantum mechanics (QM) MD simulation liberated the empirical description of the interatomic penitential in classical MD and took the role of femtosecond laser pulse excitation of electrons into account [20,26,27]. Nevertheless, the size of simulated system in *ab initio* quantum mechanics MD is limited in tens to hundreds of atoms, due to the expensive computational cost[28,29].

Therefore, it is necessary to propose an approach that ranging from the scopes from QM to MD and TTM. In our previous works [10–12,14], the electron temperature $T_e$ dependent $C_e$, $k_e$ and $G_{e-ph}$ were modeled to study the femtosecond and picosecond laser heating of copper film. This paper applies QM-MD-TTM integrated framework [14], which determines the laser excitation dependent electron thermophysical properties $C_e$, $k_e$ and $G_{e-ph}$ from QM calculation and implements them into the TTM description of the electron subsystem. The lattice subsystem is simulated by using classical MD simulation and coupled with the TTM by using $G_{e-ph}$ related heat transfer in each MD time step. Different laser fluences are applied in the silver film to investigate the corresponding thermal response of the silver film upon femtosecond laser heating. Melting and thermal ablation induced by femtosecond laser pulse heating of silver film, are observed from the simulation results. The approaches in electron scale, atomic scale and continuum scale are integrated to construct the QM-MD-TTM multiscale framework in this paper. They provide essential information to each other and capture the details of femtosecond laser interaction with silver film from their unique perspectives.

## 2. Modeling and simulation

### 2.1 QM determination of electron thermophysical parameters

QM determination can provide electron thermophysical parameters to the combined MD-TTM simulation, which includes the detailed changes of electron density of states $g$, Fermi-Dirac distribution function $f$, electron-phonon spectral function $\alpha^2 F(\Omega)$. The primary objective is to quantitatively determine the dependence of thermophysical parameters $C_e$, $k_e$ and $G_{e-ph}$ when the thermal excitation of electron is below the Fermi energy $\varepsilon_F = 5.49 \, eV$ of silver [30].

As seen in [30], the internal energy of electron subsystem $E_e$ is express as $E_e = (1/V_c) \int_{-\infty}^{\infty} g f \varepsilon d\varepsilon$, where $V_c$ is the volume of the silver unit cell and $\varepsilon$ is energy level of the electron subsystem per unit cell. The derivative of $E_e$ with respect to electron temperature $T_e$ is electron heat capacity $C_e$ [11]. Since both $g$ and $f$ are $T_e$ dependent variables, $C_e$ is expressed in terms of the following equation at given $T_e$

$$C_e|_{T_e} = \frac{1}{V_c} \int_{-\infty}^{\infty} (\frac{\partial g|_{T_e}}{\partial T_e} f|_{T_e} + g|_{T_e} \frac{\partial f|_{T_e}}{\partial T_e}) \varepsilon d\varepsilon, \qquad (1)$$

where $f$ is equal to $\{exp[(\varepsilon - \mu)/k_B T_e] + 1\}^{-1}$ and $k_B$ is the Boltzmann constant $1.3804 \times 10^{-23} J/K$. The chemical potential $\mu$ in $f$ also varies with $T_e$. When $T_e \ll T_F = 6.38 \times 10^4 \ K$ for silver [30], the excited electron is around the Fermi surface, namely $5s^1$ electron. Then $C_e$ can be expressed as a linear function of $T_e$, namely, $C_e = \gamma T_e$. According to the free electron gas model, $\gamma$ is defined as $\pi^2 n_e k_B^2/(2\varepsilon_F)$. When $T_e$ becomes higher, the lower band electrons $4d^{10}$ will be excited, which leads to the breakdown of $C_e = \gamma T_e$.

The thermal excitation of electron also induces the change of electron thermal conductivity, $k_e$. When $T_e \ll T_F$, $k_e$ is calculated from the Drude model [13] by including $C_e$ from Eq. (1), i.e.,

$$k_e|_{T_e} = \frac{1}{3V_c} v_F^2 \tau_e|_{T_e} \int_{-\infty}^{\infty} (\frac{\partial g|_{T_e}}{\partial T_e} f|_{T_e} + g|_{T_e} \frac{\partial f|_{T_e}}{\partial T_e}) \varepsilon d\varepsilon, \qquad (2)$$

where $v_F = 1.39 \times 10^6 \ m/s$ is the Fermi velocity of silver [30]. Since both electron-electron and electron-phonon scatterings contribute to the electron collision frequency, the total electron scattering time of $\tau_e$ is computed from the reciprocal of the summation of electron-electron scattering rate $\tau_{e-e}^{-1} = AT_e^2$ and electron-phonon scattering rate $\tau_{e-ph}^{-1} = BT_l$, i.e., $\tau_e = (AT_e^2 + BT_l)^{-1}$, where the parameters $A = 3.57 \times 10^6 \ s^{-1}K^{-2}$ and $B = 1.12 \times 10^{11} \ s^{-1}K^{-1}$ are two material constants of silver[30]. When $T_e$ is in the order of tens of thousands degrees Kevin and $T_l$ is in the order of several hundreds of degrees Kevin, $\tau_e$ is dominated by $T_e$. It indicates that at the initial point of femtosecond laser excites electrons, the heat conduction of electron subsystem is not affected by the thermalization of the lattice subsystem.

When it comes to the effective electron-phonon coupling factor $G_{e-ph}$, a considerable number of studies were conducted [15,31,32]. Deriving from the expression of thermal energy transport in electron-phonon collision equations, a free electron gas model was established and described $G_{e-ph}$ as $\pi^2 m_e n_e v_s^2/(6\tau_e T_e)$ by Kaganov *et al.* [31], where $m_e$ denotes the effective mass of electrons and $v_s$ is the speed of sound. In addition, as pointed out by Anisimov *et al.* [15], the expression of $\pi^2 m_e n_e v_s^2/(6\tau_e T_e)$ becomes a constant when $T_e \approx T_l \gg T_D$ (Debye temperature of silver $221 \ K$). Chen *et al.* [32] deduced a phenomenological model by evaluating $\tau_e$ and including both $T_e$ and $T_l$ in calculating $G_{e-ph}$ as $G_{RT}[A/B(T_e + T_l) + 1]$, where $G_{RT}$ was coupling factor at room temperature. Nevertheless, all the aforementioned derivation approaches of $G_{e-ph}$ are from empirical estimation, a full QM modeling of $G_{e-ph}$ is to be developed in this paper.

By recalling the definition of $G_{e-ph} = (\partial E_{e-ph}/\partial t)/[(T_e - T_l)V_c]$, the most important task is to obtain $\partial E_{e-ph}/\partial t$ through QM modeling. When electrons are excited, the variations of $g(\varepsilon)$, $f(\varepsilon)$ and $\alpha^2 F(\Omega)$ contribute to the heat transfer from electron subsystem to the lattice subsystem. By taking the electron-phonon collision into account, $\partial E_{e-ph}/\partial t$ can be obtained as [33]

$$\frac{\partial E_{e-ph}}{\partial t} = \frac{4\pi}{\hbar}\sum_{k,k'}\hbar\omega_Q|M_{k,k'}|^2 S(k,k')\delta(\varepsilon_k - \varepsilon_{k'} + \hbar\omega_Q), \tag{3}$$

where $\hbar = 1.054 \times 10^{-34} Js$ is the reduced Planck constant, $\omega_Q$ is the phonon frequency at the phonon quantum number $Q$, and $k$ and $k'$ denote the electron quantum number at initial and final states, respectively. The scattering probabilities of electrons at initial energy $\varepsilon_k$ and final energy $\varepsilon_{k'}$ are described by the matrix $M_{k,k'}$. $S(k,k')$ is equal to $[(f_k - f_{k'})n_Q - f_{k'}(1-f_k)]$, which is defined as the thermal factor to characterize the phonon absorption and emission during electron-phonon scattering. $n_Q$ is the Bose-Einstein distribution of phonons, $[exp(\hbar\omega_Q/k_B T_l) - 1]^{-1}$.

By introducing the electron-phonon spectral function $\alpha^2 F(\varepsilon, \varepsilon', \Omega)$ at high $T_e$, $2(\hbar g(\varepsilon_F)|_{T_e})^{-1}\sum_{k,k'}|M_{k,k'}|^2|_{T_e}\delta(\omega_Q - \Omega)\delta(\varepsilon_k - \varepsilon)\delta(\varepsilon_{k'} - \varepsilon')$ and multiplying the three integrals, $\int_{-\infty}^{\infty}\delta(\varepsilon_k - \varepsilon)d\varepsilon$, $\int_{-\infty}^{\infty}\delta(\varepsilon_{k'} - \varepsilon')d\varepsilon'$ and $\int_0^{\infty}\delta(\omega_Q - \Omega)d\Omega$, $G_{e-ph}$ at given $T_e$ becomes

$$G_{e-ph}|_{T_e} = \frac{1}{V_c}\frac{2\pi g(\varepsilon_F)|_{T_e}}{T_e - T_l}\{\int_0^{\infty}[\int_{-\infty}^{\infty}(\int_{-\infty}^{\infty}\alpha^2 F(\varepsilon, \varepsilon', \Omega)|_{T_e}S(\varepsilon, \varepsilon')|_{T_e}\delta(\varepsilon - \varepsilon' + \hbar\Omega)d\varepsilon)d\varepsilon']\hbar\Omega d\Omega\}. \tag{4}$$

The energy conservation requires that $\varepsilon' - \varepsilon = \hbar\omega$, and the electron-phonon spectral function is approximated as $\alpha^2 F(\varepsilon, \varepsilon', \Omega)|_{T_e} = g(\varepsilon)|_{T_e}g(\varepsilon + \hbar\Omega)|_{T_e}\alpha^2 F(\varepsilon_F, \varepsilon_F, \omega)/[g(\varepsilon_F)|_{T_e}]^2$ [34], where $\alpha^2 F(\varepsilon_F, \varepsilon_F, \omega)$ is the electron-phonon spectral function at the Fermi energy level $\varepsilon_F$. Moreover, at the limit of $k_B T_e \gg \hbar\omega$ and $k_B T_l \gg \hbar\omega$, the thermal factor becomes $S(\varepsilon, \varepsilon')|_{T_e} = [f(\varepsilon) - f(\varepsilon + \hbar\Omega)](T_e - T_l)k_B/(\hbar\Omega)$. Because the energy range of electrons is much wider than that of phonons, $g(\varepsilon)$ is approximately equal to $g(\varepsilon + \hbar\Omega)$, and $[f(\varepsilon) - f(\varepsilon + \hbar\Omega)]/(\hbar\Omega)$ is rewritten as $-\partial f/\partial\varepsilon$. In addition, at the high $T_e$ limit, the second moment of $\alpha^2 F(\Omega)|_{T_e}$ is simplified as $\lambda\langle\omega^2\rangle|_{T_e} = 2\int_0^{\infty}\alpha^2 F(\Omega)|_{T_e}\Omega d\Omega$ [33]. Therefore, Eq. (4) becomes

$$G_{e-ph}|_{T_e} = \frac{1}{V_c}\frac{\pi\hbar k_B\lambda\langle\omega^2\rangle|_{T_e}}{g(\varepsilon_F)|_{T_e}}\int_{-\infty}^{\infty}g|_{T_e}^2(-\frac{\partial f|_{T_e}}{\partial\varepsilon})d\varepsilon, \tag{5}$$

In Eqs. (1), (2) and (5), $g|_{T_e}$, $f|_{T_e}$, $g(\varepsilon_F)|_{T_e}$ and $\lambda\langle\omega^2\rangle|_{T_e}$ are all $T_e$ dependent variables, which can be calculated by *ab initio* QM approach. At low $T_e$, because $\partial f|_{T_e}/\partial\varepsilon$ is a delta function, $G_{e-ph}|_{T_e}$ in Eq. (5) becomes $\pi\hbar k_B\lambda\langle\omega^2\rangle|_{T_e}g(\varepsilon_F)|_{T_e}/V_c$. The derivation of Eq. (5) was reported without considering the laser induced changes of $\lambda\langle\omega^2\rangle$, $g(\varepsilon_F)$ and $g$ [8]. In this paper, we obtain $g|_{T_e}$, $f|_{T_e}$, $g(\varepsilon_F)|_{T_e}$ and $\lambda\langle\omega^2\rangle|_{T_e}$ at given $T_e$ as the first step [10] and make connections of each individual $C_e|_{T_e}$, $k_e|_{T_e}$ and $G_{e-ph}|_{T_e}$ to get the $T_e$ dependent thermophysical parameters of laser excited electrons.

## 2.2 Combined MD-TTM simulation and its integration with QM

Due to the limitation of TTM in probing the atomistic scale details of lattice temperature $T_l$ evolution and mechanisms of phase change phenomena induced by laser heating, MD simulation is introduced to model the motion of atoms in lattice subsystem. Meanwhile, the electron subsystem is characterized by TTM to describe the laser energy deposition, electron thermalization, heat conduction, and the part of thermal energy transferring from the electron subsystem to the lattice subsystem. The electron subsystem is divided into $N$ cells with $N_V$ (variable) atoms in the volume of $V_N$. In this paper, the $T_e$ dependent electron thermophysical parameters $k_e$, $C_e$ and $G_{e-ph}$ in Eqs. (1), (2) and (5) are implemented into the energy equation of electron subsystem in TTM

$$C_e \frac{\partial T_e}{\partial t} = \frac{\partial}{\partial x}\left(k_e \frac{\partial T_e}{\partial x}\right) - G_{e-ph}(T_e - T_l) + S(x,t), \qquad (6)$$

where $S(x,t)$ is the source term due to laser heating. Equation (6) was solved by a finite difference method (FDM) as a 1D problem in $x$-direction. The temporal evolution of $S(x,t)$ obeys Gaussian distribution,

$$S(x,t) = \frac{0.94 J_{abs}}{t_p L_{op}} exp\left(-\frac{x}{L_{op}+L_{ba}}\right) exp\left[-2.77 \frac{(t-t_0)^2}{t_p^2}\right], \qquad (7)$$

where $J_{abs}$ is the absorbed laser fluence. The attenuation of absorbed laser energy agrees with the Beer-Lambert law. The optical penetration depth $L_{op} = 12\ nm$ [35] and ballistic laser energy transport depth $L_{ba} = 56\ nm$ [30] are used to represent the effective range of laser energy deposition $L_p$. The temporal center point of the laser beam locates at $t_0 = 25\ ps$, which indicates the maximum value of laser intensity occurs at $25\ ps$. The full width at half maximum (FWHM) is $500\ fs$, which is defined as the laser pulse duration $t_p$.

In one MD time step $\Delta t_{MD} = 1 fs$ of simulating the atomic motion in lattice subsystem, the FDM computation of electron subsystem is performed $n_t = 200$ times of with a FDM time step of $\Delta_{FDM} = 0.005 fs$. The von Neumann stability criterion [36] has to be met, namely,

$$\Delta t_{FDM} = \frac{\Delta t_{MD}}{n_t} < 0.5 \Delta x_{FDM}^2 \frac{C_e}{k_e} = 1.5 \frac{\Delta x_{FDM}^2}{v_F^2 \tau_e}, \qquad (8)$$

At each time step of MD simulation, the energy transferring from electron subsystem to lattice subsystem for a given FDM cell is modeled as [22]

$$E_{Transfer} = \frac{\Delta t_{MD}}{n_t} \sum_{k=1}^{n_t} G_{e-ph} V_N (T_e^k - T_l), \qquad (9)$$

where $T_e^k$ is the average electron temperature in each $\Delta t_{MD}$. Therefore, the equation of atomic motion include the force from interatomic potential and the force induced by thermal energy transferred to the lattice subsystem, i.e.,

$$m_i \frac{d^2 r_i}{dt^2} = -\nabla U + \frac{E_{Transfer}}{\Delta t_{MD}} \frac{m_i v_i^T}{\sum_{j=1}^{N_V} m_j (v_j^T)^2}, \qquad (10)$$

where $m_i$, $r_i$ and $v_i^T$ are the mass, position and thermal velocity of atom $i$. The interatomic potential $U$ of silver is described by the embedded atom method (EAM), which was developed by fitting the potential energy surface from QM calculation [37].

**2.3 Simulation details**

The QM-MD-TTM integrated framework is constructed by combing Eqs. (1), (2), (5), (6) and (10). The simulation code is developed as an extension of the TTM part in the IMD [38] and the ABINIT [39]. The QM calculation was performed by using finite temperature density function theory (FT-DFT) for perfect silver lattice for $T_e$ ranging from $300\ K$ to $3\times 10^4\ K$. For the reason that the energy separating core electrons to the valence band is tens of $eV$ (for $T_e$ in the magnitude of $10^5\ K$), the valence electrons $4d^{10}5s^1$ were taken explicitly during femtosecond laser excitation of electron. In other words, the excitations from core electrons into the valence band were not taken into account. The local density approximation (LDA) was employed in calculating the exchange and correlation energy. After convergence test, a plane wave cutoff of $28\ eV$ and a Monkhorst Pack mesh of $10\times 10\times 10$ $k$-point were adopted in the FT-DFT calculation. Fifty bands were set to allow sufficient states to be occupied of during the high laser energy excitation of electrons.

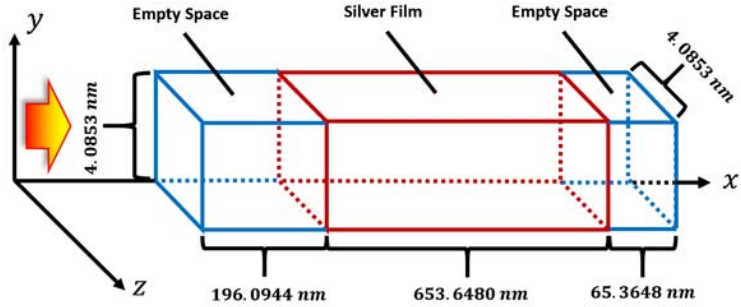

**Fig. 1 Schematic show of the established system and direction of laser irradiation.**

Figure 1 shows a schematic view of simulation system, which is consisted with three components along the incidence of laser pulse ($x$-direction). The first and third components are two empty spaces, which are set to allow thermal expansion and ablation of the silver film under femtosecond laser heating. The middle component is the silver film with equal width and length of $4.0853\ nm$ in $y$- and $z$-directions. The film thickness is $653.6480\ nm$. During the simulation, periodic boundary conditions (PBC) were applied in both the $y$- and $z$-directions of the modeled system. The front surface and rear surface of the silver film were set as free boundaries. The total simulation time was $160\ ps$, which included the first $5\ ps$ of canonical ensemble (NVT) simulation to initialize the system at equilibrium state and another $5\ ps$ of micro-canonical ensemble (NVE) simulation to check whether the thermal equilibrium had reached. Since the QM part occupied major computational load in the QM-MD-TTM integrated simulation, the QM part was carried out firstly to achieve the quantitative relationship (by means of polynomial fitting) of electron temperature with electron heat capacity, electron thermal conductivity (which also depends on the lattice temperature) and electron-phonon coupling factor. Subsequently, the MD-TTM combined simulation was carried out by calling the polynomial functions.

# 3. Results and discussion

## 3.1 Electron thermophysical parameters $C_e$, $k_e$ and $G_{e-ph}$

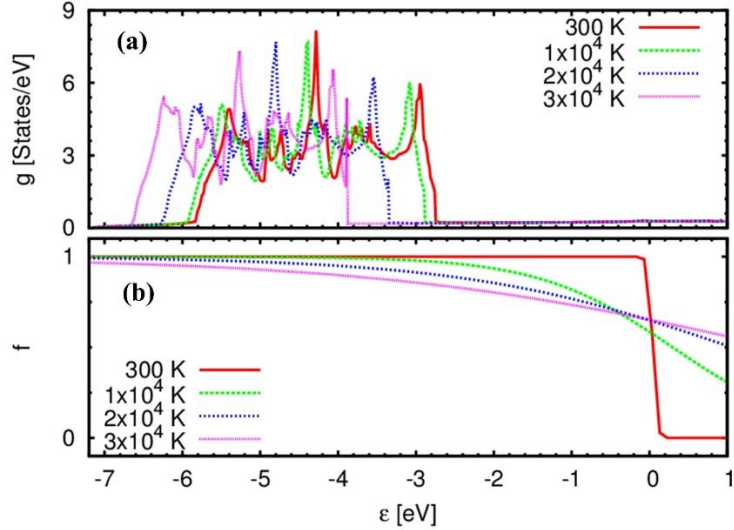

**Fig. 2** (a) The electron density of states $g$, (b) Fermi-Dirac distribution $f$ obtained from *ab initio* quantum mechanical calculation.

Before calculating the electron heat capacity $C_e$, electron thermal conductivity $k_e$ and effective electron-phonon coupling factor $G_{e-ph}$ modeled in Section 2.1, electron density of states $g$ and Fermi-Dirac distribution function $f$ have to be available from QM calculation. Figure 2 shows the calculated $g$ and $f$ of $T_e$ at 300 $K$, $1 \times 10^4$ $K$, $2 \times 10^4$ $K$, and $3 \times 10^4$ $K$. The zero point of $\varepsilon$ is set as the Fermi energy $\varepsilon_F$. As seen in Fig. 2(a), $g$ shows significant changes of electron for $T_e$ ranging from $1 \times 10^4$ $K$ to $3 \times 10^4$ $K$. Whereas, the variation of $g$ at $T_e < 1 \times 10^4$ $K$ is marginal, which demonstrated the large scale excitation of $d$ band electrons starts at $T_e \geq 10,000$ $K$. With the increase of $T_e$, $g$ exhibits a gradual shrinkage and left shift to the lower $\varepsilon$. The change of $g$ was explained by the changes of electronic screening [40]. Owing to the laser excitation of the electrons, the number of $4d^{10}$ decreases, which results in a more attractive electron-ion potential. Therefore, the overall distribution of electron states move to lower $\varepsilon$. Figure 2(b) indicates that $f$ gets smeared at higher $T_e$. In addition, the central value of $f = 0.5$ locates at higher $\varepsilon$ for the case of higher $T_e$, which is sourced from the increase of $\mu$ at excited electron states. The detailed values of $\mu - \varepsilon_F$ at given $T_e$ is showed in Fig. 3(a). It can be seen that the increase of $\mu$ becomes obvious for $T_e$ from $1 \times 10^4$ $K$ to $3 \times 10^4$ $K$. When the $T_e$ approaches to the Fermi temperature of silver, the increase of $\mu$ becomes slower. According to [30], the number of valence electrons is $N_e^v = \int_{-\infty}^{\infty} f g d\varepsilon$. Therefore, in order to keep the number of valence electrons as constant, the left shift and shrinkage of $g$ has to be compensated by the overall right movement of $f$ (including the increasing $\mu$).

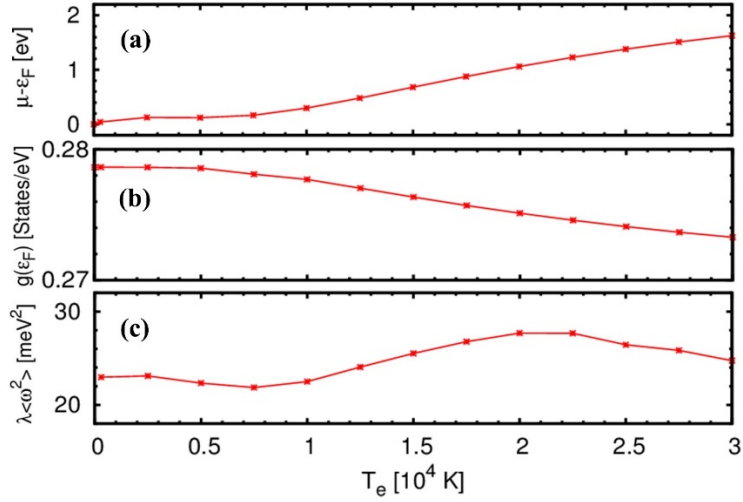

**Fig. 3** Evolutions of (a) the difference of chemical potential $\mu$ from Fermi energy $\varepsilon_F$; (b) the electron density of states $g$ at Fermi energy $\varepsilon_F$; (c) the second momentum of $\lambda\langle\omega^2\rangle$ of electron-phonon spectral function $\alpha^2 F(\Omega)$ at given $T_e$.

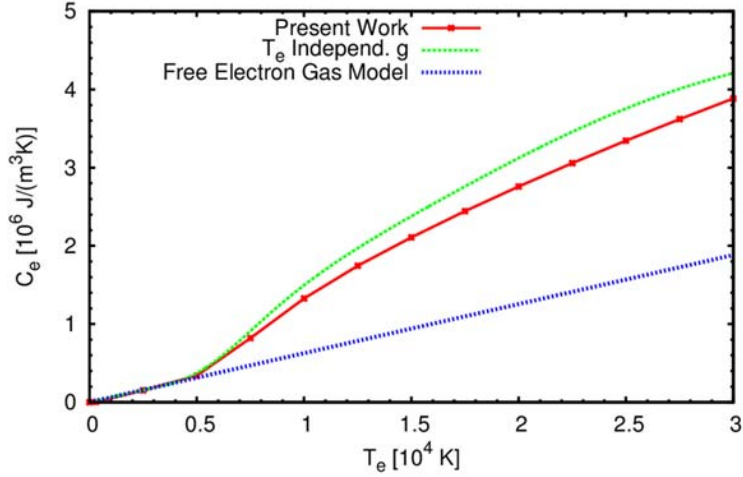

**Fig. 4** The electron temperature $T_e$ dependent electron heat capacity $C_e$: (a) by taking the variations of density of states $g$ and Fermi-Dirac distribution $f$ at different $T_e$; (b) by ignoring the change of $g$ at different $T_e$ [8]; (c) by estimating from free electron gas model [41].

By including the effects of $T_e$ induced variations of $g$ and $f$ in Eq. (1), the calculated electron heat capacity $C_e$ at given $T_e$ is plotted in Fig. 4. The free electron number density of silver is $5.57 \times 10^{28}/m^3$, which takes four electron $5s^1$ per face centered unit (FCC) unit cell. The fermi energy $\varepsilon_F$ of silver is $5.49\ eV$ [41]. Therefore, the theoretical $\gamma_{th}$ derived from Section 2.1 is $62.76\ J/(m^3 K^2)$, which is approximately equal to the experimentally obtained coefficient $63.30\ J/(m^3 K^2)$ [42]. For the purpose of comparison, another QM calculated $C_e$ without including the effect of $T_e$ induced variation of $g$ [8], and $C_e = \gamma_{th} T_e$ obtained from the free electron gas model are also plotted in Fig. 4. When $T_e < 0.5 \times 10^4\ K$, there are overlaps among three $C_e$ computed by using different methods, which implicitly demonstrates the number of excited electron is only $5s^1$. When $T_e$ continues increasing from $0.5 \times 10^4~K$ to higher values, the $4d^{10}$ electrons get excited as a result of breakdown of the linear relationship between $C_e$ and $T_e$, as well as the redistribution of $g$. Due to the left shift of $g$ to lower $\varepsilon$ and narrowing the $d$-band $g$ distribution, the term $(\partial g|_{T_e}/\partial T_e)f|_{T_e}$ in Eq. (1) brings negative effect in calculating $C_e$, which leads to smaller $C_e$ in the present work than that treating $g$ as a $T_e$ independent parameter [8].

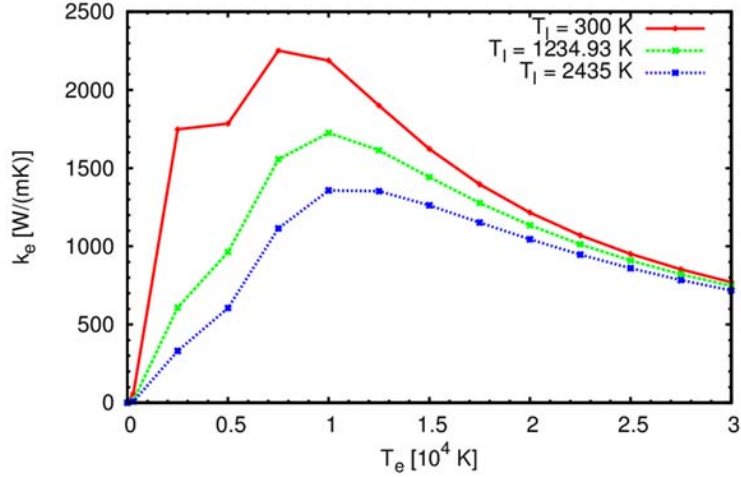

**Fig. 5** The electron temperature $T_e$ dependent electron thermal conductivity $k_e$ at lattice temperatures of room temperature (300 K), melting point ($T_m = 1234.93~K$) and boiling point $T_b = 2435~K$.

Figure 5 shows the electron thermal conductivity $k_e$ calculated from Drude model in Eq. (2). Since $k_e$ is a function of both $T_e$ and $T_l$, there are three cases (for $T_l = 300~K$, $1234.93~K$ and $2435~K$) plotted in Fig. 5; these cases correspond to the conditions of $T_l$ at room temperature, melting point and boiling point. When the electron are heated at the same degree, higher $T_l$ results in smaller $k_e$. When $T_e < 0.75 \times 10^4~K$, even though the increase of $T_e$ brings negative effect to the right side of Eq. (2), the increase of $C_e$ plays the dominant role in promoting the electron heat conduction, which leads to the monotonous increase of $k_e$. As seen in Fig. 4, the increasing rate of $C_e$ from $0.5 \times 10^4~K$ to $0.75 \times 10^4~K$ is faster than that from $300~K$ to $0.5 \times 10^4~K$. Nevertheless, for the case of $T_l = 300~K$, due to the quadratic increase of $T_e^2$ and relatively smaller $T_l$ than the other two cases in Fig. 5, $k_e$ (for $T_l = 300~K$) shows the most complex increasing curve among the three cases. With the continuous increase of $T_e$ above $1 \times 10^4~K$, $k_e$ for all the three cases render decreasing trends. It should be noted that the differences among $k_e$ for the three cases become smaller at higher $T_e$.

Before obtaining the effective electron-phonon coupling factor $G_{e-ph}$, the $T_e$ dependent $g(\varepsilon_F)$ and $\lambda\langle\omega^2\rangle$ in Eq. (5) have to be available as prerequisites. Figure 3(b) and 3(c) show $g(\varepsilon_F)$ and $\lambda\langle\omega^2\rangle$ at given $T_e$. As seen in Fig. 3(b), when $T_e > 0.5 \times 10^4$, $g(\varepsilon_F)$ continuously decreases with increasing $T_e$. Recalling $g(\varepsilon_F)$ is the denominator in Eq. (5), it can be concluded that decrease of electron density of state at Fermi level contributes to enhancing $G_{e-ph}$ at higher $T_e$. The gentle change of $g(\varepsilon_F)$ from $300~K$ to $0.5 \times 10^4$ agrees with the aforementioned $g$ in computing $T_e$ dependent $C_e$. Whereas, the change of $g(\varepsilon_F)$ at different levels of electron excitation is absent in calculation of $G_{e-ph}$ in [8]. On the basis of computing the electron-phonon spectral function $\alpha^2 F(\Omega)$, its second moment $\lambda\langle\omega^2\rangle$ is obtained and plotted in Fig. 3(c). According to the reported electron-phonon coupling constant $\lambda = 0.12$ in [43], $\lambda\langle\omega^2\rangle$ is $22.5\ meV^2$, which reveals that the calculated $\lambda\langle\omega^2\rangle$ at low $T_e$ is in good agreement of with the empirical value. However, when $T_e$ increases from $0.75 \times 10^4 K$ to $2 \times 10^4\ K$, obvious increment of $\lambda\langle\omega^2\rangle$ is exhibited in Fig. 3(c), which indicates that the strength of electron-phonon coupling evolves stronger than stronger. The calculated results of $\alpha^2 F(\Omega)$ in this paper show that when $T_e > 2 \times 10^4\ K$, the high frequency $\Omega$ parts of $\alpha^2 F(\Omega)$ gradually decrease. Therefore, $\lambda\langle\omega^2\rangle$ becomes smaller with the increase of $T_e$ above $2 \times 10^4\ K$. The overall evolution of $G_{e-ph}$ is depicted in Fig. 6, which presents increasing trend with the increase of $T_e$. When $0.25 \times 10^4\ K < T_e < 0.5 \times 10^4\ K$, there are marginal changes of $g(\varepsilon_F)$ and slight decrease of $\lambda\langle\omega^2\rangle$. However, $G_{e-ph}$ still shows small increase, which is caused by the dominant role of Fermi smearing. For the purpose of comparison, another *ab initio* calculated $G_{e-ph}$ by treating $g(\varepsilon_F)$, $g$ and $\lambda\langle\omega^2\rangle$ as constants [8], and Chen's phenomenological $G_{e-ph}$ [32] are plotted in Fig. 6. The parameter $G_{RT}$ in the Chen's phenomenological model was taken as $3.5 \times 10^{16}\ W/(m^3 K)$ [44], which was measured for a silver film with thickness of $45\ nm$ in femtosecond by using optical transient-reflection technique. Significant discrepancies are shown between $G_{e-ph}$ calculated in this paper and $G_{e-ph}$ reported in [8]. The treatments of $T_e$ independent $g(\varepsilon_F)$, $g$ and $\lambda\langle\omega^2\rangle$ eliminate the impacts of these three parameters to the electron-phonon coupled heat transfer at specified $T_e$ [8], which results in deviations from $G_{e-ph}$ determined in the present work. The phenomenological $G_{e-ph}$ includes both the effects brought by $T_e$ and $T_l$, but it is limited by the precondition of experimentally measured $G_{RT}$. In the QM-MD-TTM integrated simulation, the $T_e$ dependent $C_e$ and $G_{e-ph}$ were fitted as polynomial functions, which enabled the prediction of coupling factor and electron heat capacity when electron temperature is above $30,000\ K$.

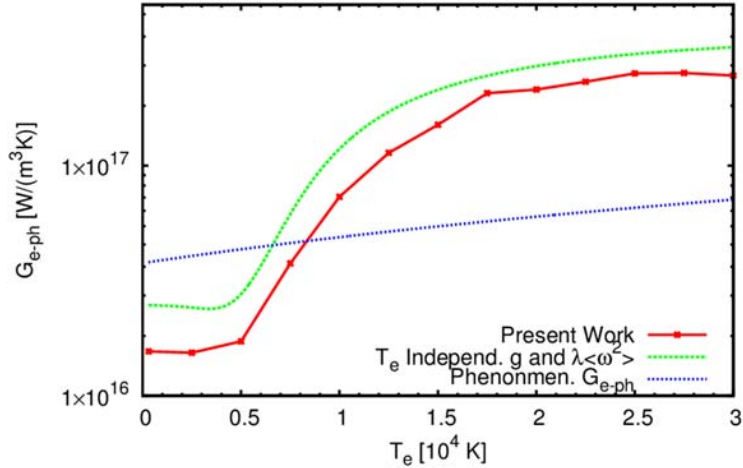

**Fig. 6** The electron temperature $T_e$ dependent effective electron-phonon coupling factor $G_{e-ph}$ from *ab initio* quantum mechanical calculation (a) by including the $T_e$ dependence of $g$, $f$ and $\lambda\langle\omega^2\rangle$; (b) by treating $g$ and $\lambda\langle\omega^2\rangle$ as constants [8]; (c) from Chen's phenomenological model [32].

## 3.2 Laser melting of the silver film

By setting the absorbed laser fluence as $J_{abs}$ as $0.1\ J/cm^2$, both heterogeneous melting and homogeneous melting of the silver film were found. When it comes to study the phase transition from solid to liquid, to identify solid and liquid states is a challenging problem. There are several approaches available to differentiate the solid and liquid states, such as, setting up a criterion from local order calculation [22], comparing the lattice temperature with the melting point [45] and the measured mean square displacement with Lindemann's criterion [46], performing common neighbor analysis (CNA) method to characterize the deformation of FCC silver crystal [47], as well as calculating the spatial density distribution to find the homogenous melted region [11].

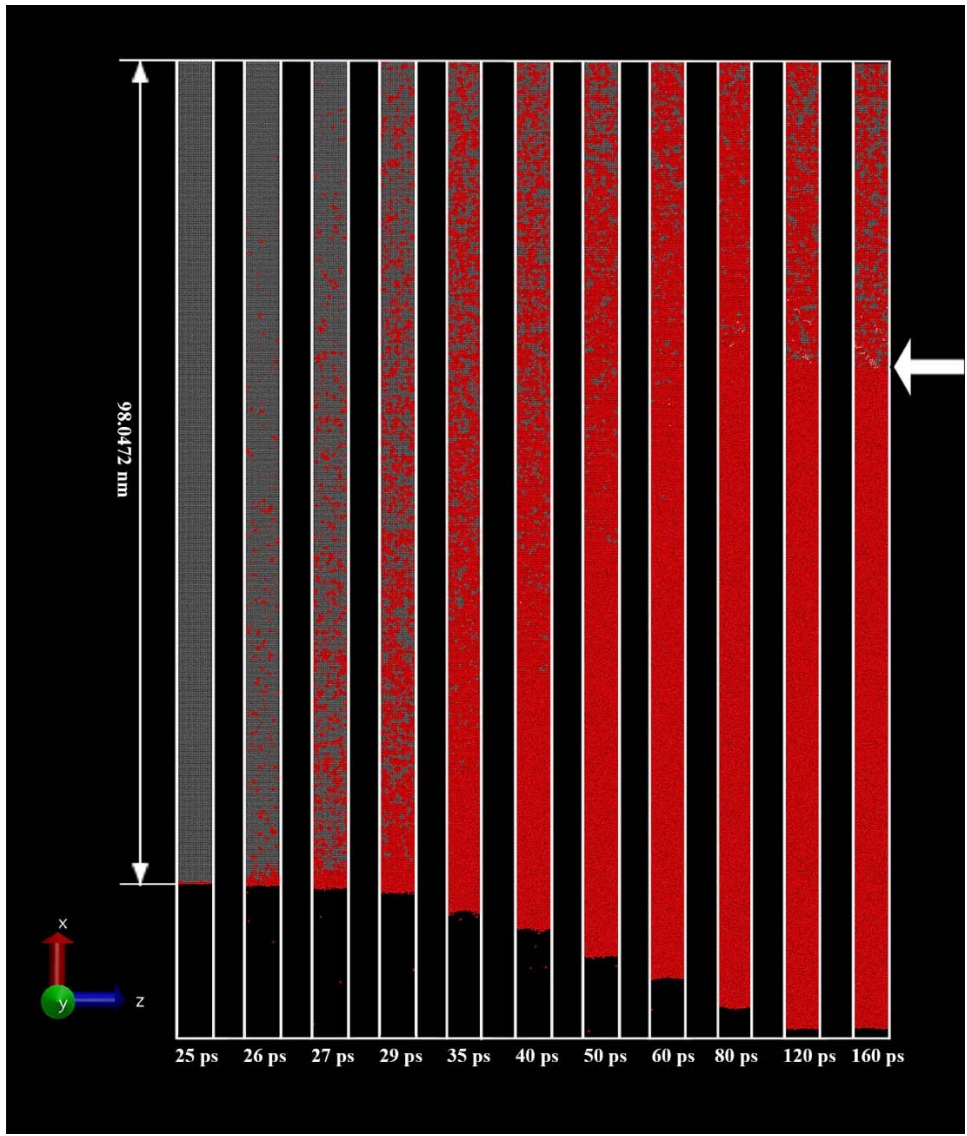

**Fig. 7** Snapshots of atomic view for the front surface from $25\ ps$ to $39\ ps$. The colors are characterized from common neighbor analysis (CNA). Grey color denotes face centered cubic (FCC) lattice structure of solid silver. Red color is for those laser melted regions. (Note: Color online for this figure.)

In this paper, CNA was used to study the local crystal structure change of silver film. At each MD time step, numbers representing the crystal structure were output along with the atomic coordinate. The melting process was detected by rendering overall atomic CNA values with given colors. For the FCC silver crystal, it was rendered in light (grey) color. The laser melted silver (disordered atomic configuration) was plotted in dark (red) color. By visualizing from OVITO [48], the snapshots of atomic configurations with colors labeled from CNA results are shown in Fig. 7. Only part of the silver film is shown. With the evolution of simulation time, the number of disordered atoms are increasing. It can be seen in Fig. 7 that the heterogeneous melting penetrates to tens of nanometers into the silver film in $1\ ps$ (from $25\ ps$ to $26\ ps$). Meanwhile, there is a small region of pure red in the front surface of the silver film, which demonstrates the homogenous melting appears. Owing to nonequilibrium state between the electron subsystem and the lattice subsystem, thermal energy continuously transports from the hot electron to cold lattice, which results in some of the homogenous melting regions are heated into homogenous regions. The interface between homogenous melting and heterogeneous melting are identified by the whether all the silver atoms are melted. As seen in the homogenous melting region in Fig. 7, there is no silver atoms aggregating in terms of face centered cubic (FCC) state. Whereas in the heterogeneous melting region (deeper below the homogenous melting), there are mixed silver atoms aggregating in terms FCC and non-FFC states. As seen from $26\ ps$ to $60\ ps$, with larger number of regular FCC arranged silver atoms being heated, the heterogeneous melting region develops deeper inside the silver film. Some individual atoms locating in the front surface of silver film is heated and vaporized into the empty space. Between the totally disordered atoms and the partially regular FCC arranged atoms, a rough interface of homogenous melting and heterogeneous melting is seen at $120\ ps$. From $120\ ps$ to $160\ ps$, $x$-direction advancement of the interface stops, which demonstrates the femtosecond laser induced thermal melting of silver film has been fully developed. By employing femtosecond electron diffraction, structural evolution of an aluminum film reported the resemble long range disorder and short range liquid structure [49].

The temporal and spatial distribution of normalized density ($\rho^* = \rho/\rho_0$, where $\rho_0 = 10.49\ g/cm^3$ is density of silver at room temperature) is seen in Fig. 8(a). It can be seen that since the central point ($25\ ps$) of femtosecond laser heating, a compressed region ($\rho^* > 1$) is generated at the front surface of the silver film. The compressed region develops deeper inside the film and produces greater number of compressed regions with intervals of normal region ($\rho^* \approx 1$). Moreover, there are expanded regions ($\rho^* < 1$) appearing in path right after the travelling of the compressed region. It should be noted that some expanded regions are generated at the rear surface since $25\ ps$ and travel to the opposite direction of the major compressed and expanded regions. By taking an overall look of Fig. 8(a), the individual compressed and expanded regions form density waves. Both the front and rear surfaces expand with the evolution of time. On the front and rear surfaces of the silver film, $\rho^*$ shows values smaller than 0.6. For the reason that $\rho^*$ of the vaporized atoms seen in Fig. 7 are too small, they are not seen in the empty space in Fig. 8. A region with stable and uniform density lower than $\rho_0$ appears below the front surface, which agrees with the heterogeneous melting discussed from CNA result in Fig. 7.

Electron temperature $T_e$ and lattice temperature $T_l$ of the femtosecond laser heated silver film are depicted in Fig. 8(b) and 8(c), respectively. The spatial distribution of $T_e$ at $25\ ps$ shows abrupt elevation due to the absorption of ultrashort laser energy. For the region $x \geq 392.1888\ nm$, because of relatively smaller $T_e - T_l$ than that in the region $196.0944\,nm < x < 392.1888\,nm$, the electron subsystem get thermal equilibrium with the lattice subsystem $25\,ps$ to $45\,ps$. However, for the region $196.0944\,nm < x < 392.1888\,nm$, comparing $\rho^*$ distribution in Fig. 8(a) with $T_l$ distribution in Fig. 8(c), superheating is concluded (the melting point of silver $T_m = 1234.93\,K$) from $25\,ps$ to $45\,ps$. At interface between heterogeneous melting region and homogeneous melting region, the same clear interface of $T_l$ is shown between the two regions. In Fig. 8(c), when $x \geq 392.1888\,nm$, a band of atoms exhibit higher $T_l$ for the compressed region than $T_l$ of the neighboring atoms from $80\,ps$ to $160\,ps$. It is because of the atoms in the compressed region carry greater thermal energy than their neighboring atoms. Moreover, it should be noted that the present simulation only focuses on the temporal range from 25 ps (maximum laser intensity of femtosecond laser pulse irradiation) to 160 ps. With time goes on, a final thermal equilibrium state with the cooling down of the front surface and continuous heating of the rear surface of the bulk silver film will occur. From the thermophysical perspective, the femtosecond laser heated electron subsystem and electron-phonon coupled energy transfer to the lattice subsystem, result in laser melting. More specifically, the heterogeneous melting is induced by the fast heat conduction of the heated electrons from the front surface to the deeper region. However, due to the insufficient energy in the electron subsystem, partial silver atoms get melted, which is observed as heterogeneous melting. As seen in the Fig. 5 of the electron thermal conductivity, when the electron temperature is greater than 10,000 K, both higher electron and lattice temperatures lead to lower electron thermal conductivity, which further weakens the locally deposited laser energy conducts to the deeper electron regions. Therefore, all the silver atoms are melted as homogenous melting. Moreover, the greater electron heat capacity and electron-phonon coupling factor at higher electron temperature, contribute to the larger amount of thermal energy absorbed by the lattice subsystem in the homogenous melting region.

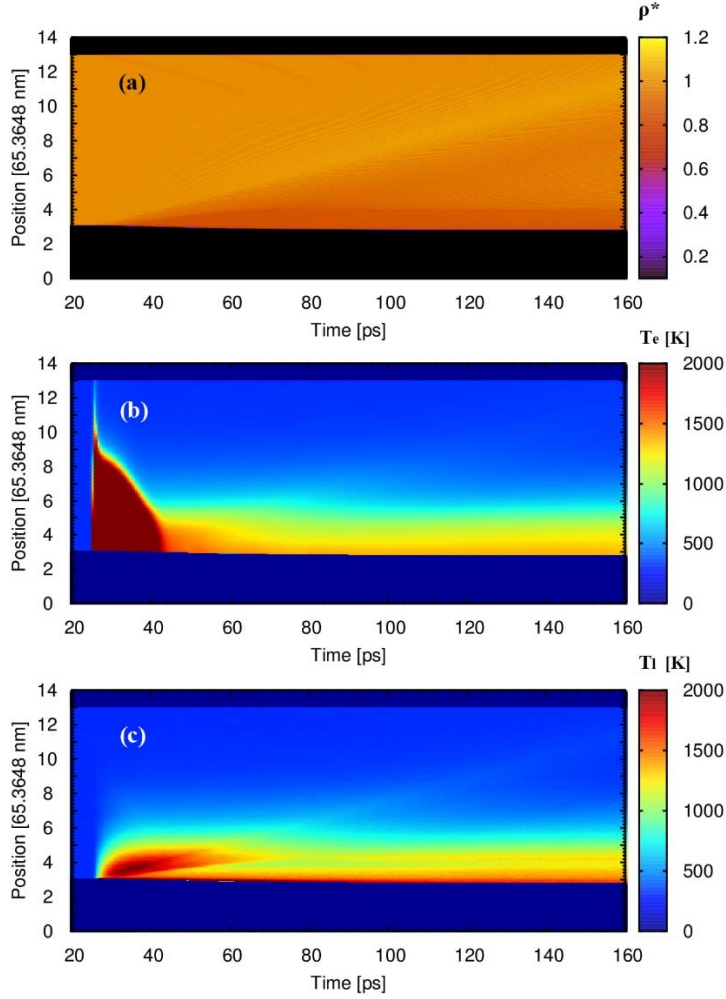

**Fig. 8** Temporal and spatial distribution of normalized density $\rho^*$, electron temperature $T_e$ and lattice temperature $T_l$ for the absorbed laser fluence $J_{abs} = 0.1\,J/cm^2$.

### 3.3 From melting to ablation

When the absorbed fluence $J_{abs}$ of incident femtosecond laser pulse increased from $0.1\ J/cm^2$ to greater value, ablation was detected from the front surface of the silver film. Figure 9 shows the simulation results of normalized density $\rho^*$, thermal stress $\tau_{xx}$ and lattice temperature $T_l$ for $J_{abs} = 0.2\ J/cm^2$.

The laser ablation occurs right below the front surface, which results from the excessively heated electron subsystem and thermal expansion of locally superheated lattice via electron-phonon coupled energy transfer. Figure 9(a) shows the temporal and spatial distribution of $\rho^*$. From $25\ ps$ to $70\ ps$, a melted region emerges from the front surface of the silver film. Unlike $\rho^*$ shown in Fig. 8(a), the melted region in Fig. 9(a) exhibits appreciable thermal expansion to the empty space and rapid melting into the inner side of the silver film. Right after $80\ ps$, the ablation happens. Atomic view of the evolution from melting to ablation is seen in Fig. 10. Owing to the increase of $J_{abs}$, the atomic snapshot at $80\ ps$ shows preliminary signs of ablation for the region from $196.0944\ nm$ to $294.1416\ nm$. A void is seen for the atoms locating right below $196.0944\ nm$. In the Fig. 10(b) at $80\ ps$, individual atoms are seen in the front space of the computational domain. Meanwhile, a large atomic aggregation tends to leave the silver film. From $80\ ps$ to $160\ ps$, the outmost atomic aggregation has the highest speed ($747\ m/s$) ablated away from the silver film, which is almost equal to the speed of thermal expansion (from $25\ ps$ to $80\ ps$) of the front surface. At $160\ ps$ in Fig. 10, there are atomic aggregations with clear boundaries ablated from the silver film. The innermost atomic aggregation is the last ablated one, which leaves a gradually enlarged distance from the bulk film. After its ablation, the homogeneous melting region present steady thickness without appreciable thermal expansion in Fig. 9(a). Comparing the location of innermost ablated aggregation Fig. 9(a) with Fig. 10(a) at $160\ ps$, it can be seen the location of the ablated aggregation is almost equal, which proves the validity of multiscale coupling between the MD simulation and TTM calculation.

In order to elucidate the mechanism behind ablation, the computed thermal stress $\tau_{xx}$ and $T_l$ are depicted in Figs. 9(b) and 9(c), respectively. Virial theorem was adopted in the calculation of $\tau_{xx}$[50]. As seen in the QM calculated results, the higher electron temperature leads to the higher electron heat capacity, electron-phonon coupling factor, and lower electron thermal conduction, which results in greater degree of locally heat electron and lattice. As seen from Fig. 9(b), high $\tau_{xx}$ (in black color) is generated since laser heating and develops into two parts at $42\ ps$. One of the high $\tau_{xx}$ develops inside the silver film, with decreasing $\tau_{xx}$ left behind its travelling path. The other high $\tau_{xx}$ is carried by the ablated atomic aggregations. To the time scale of the current simulation, the ablated atomic aggregation is metastable. Whether the atomic aggregation will maintain its shape in nanosecond and millisecond time scale is under uncertainty. From the perspective of thermal and mechanical engineering, the remained silver film after ablation of the front surface is the major interest. From $25\ ps$ to $50\ ps$, there is an obvious temperature boundary in the spatial distribution of $T_l$ in Fig. 9(c), which has temperature $\sim 2,500\ K$. Whereas, $T_l$ of the two regions below and above the boundary are up to $3,000\ K$. Comparing with the boiling point of silver $T_b = 2,435\ K$, a conclusion of locally superheated silver existing in the front surface is drawn. Due to continuous thermal energy transporting from the electron subsystem to the lattice subsystem, thermal expansion of the locally superheated region happens. Furthermore, due to the relaxation of the compressive $\tau_{xx}$ and the free boundary condition is applied in the surface of the silver film, tensile $\tau_{xx}$ is generated. Additionally, the ultrafast thermal expansion results in decrease of $T_l$ in the superheated region from $50\ ps$ to $76\ ps$ and a number of atomic aggregations being ablated subsequently. The inner ablated four atomic aggregations present smaller $T_l$ and lower $\tau_{xx}$ than the outer four atomic aggregations. In addition, Fig. 9(b) shows $\tau_{xx}$ at the starting points of the inner four ablated atomic aggregations are smaller than $\tau_{xx}$ at the deeper regions (with black contour), which indicates the ablation is driven by the excessively superheated silver and its thermal expansion. Analogical to discussion of the laser melting, homogenous melting and heterogeneous melting are observed sequentially along the $x$-direction of the remained silver film.

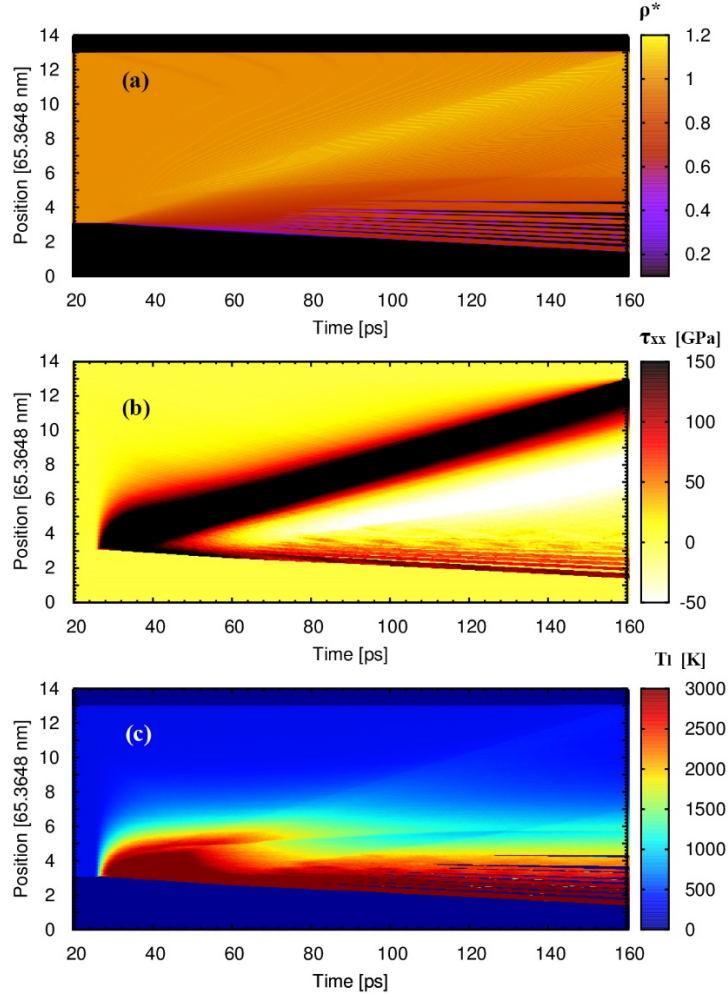

**Fig. 9** Temporal and spatial distribution of normalized density $\rho^*$, thermal stress $\tau_{xx}$ and lattice temperature $T_l$ for the absorbed laser fluence $J_{abs} = 0.2\ J/cm^2$.

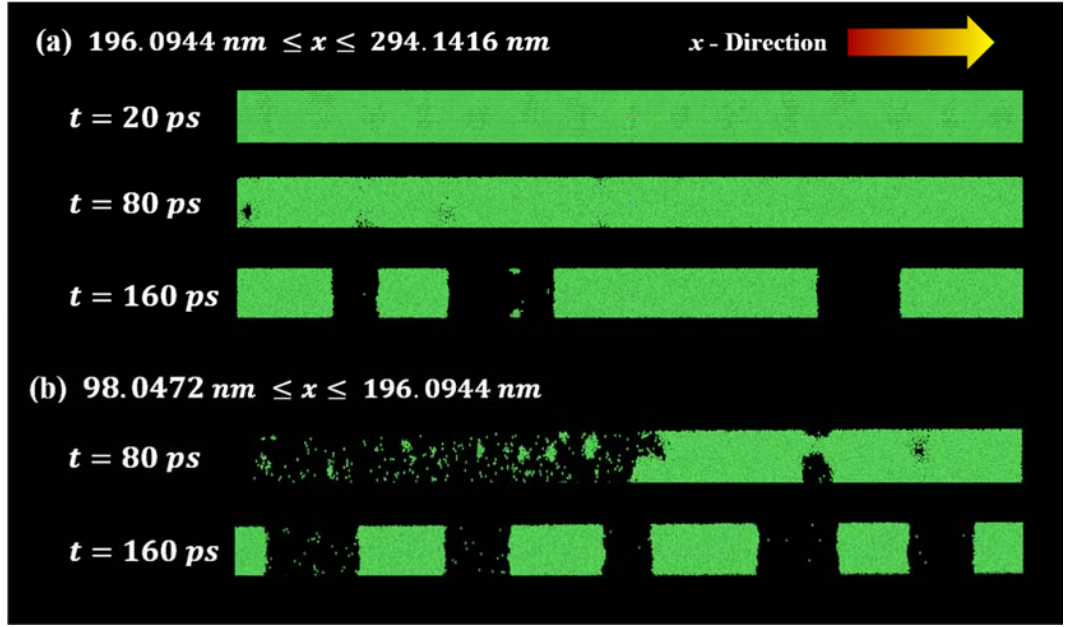

Fig. 10 Atomic view of femtosecond laser heat silver evolving from melting to ablation, (a) the atoms locating from $196.0944\ nm$ to $294.1416\ nm$; (b) the atoms locating from $98.0472\ nm$ to $196.0944\ nm$. The initial location of the front surface of the silver film is at $196.0944\ nm$. Therefore, the space from $98.0472\ nm$ to $196.0944\ nm$ is empty before laser irradiation, which is not drawn in (b).

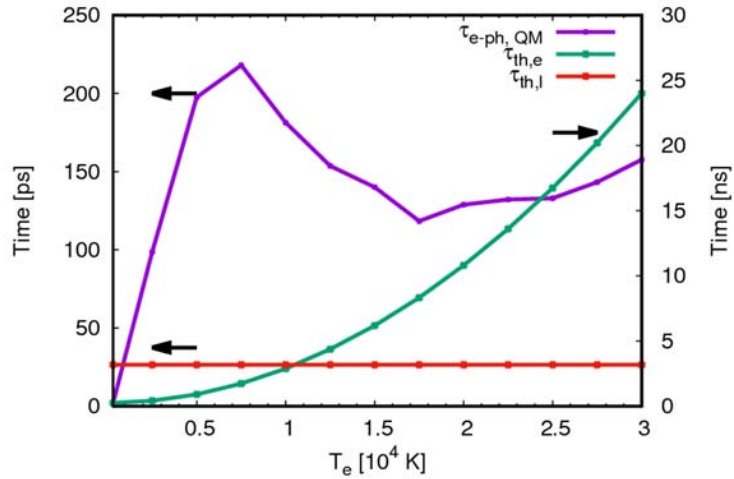

Fig. 11 $T_e$ dependent electron-phonon relaxation time $\tau_{e-ph,QM}$ estimated from QM, temporal threshold of electron thermal confinement $\tau_{th,e}$ and lattice thermal confinement $\tau_{th,l}$.

The mechanical confinement and thermal confinement are two major explanations interpreting the femtosecond laser ablation [51–53]. The confinement says that a net force exerting on the cold lattice to ablate the metal film, which obviously deviates from the simulation results of heated lattice in Fig. 9. The prerequisite of stress confinement refers the maximum timescale of laser pulse duration $t_p$ and electron-phonon scattering time $\tau_{e-ph}$ is smaller than $\tau_c = L_p/C_s =$ 25.3731 $ps$ [53], where $L_c = 68\ nm$ is the laser penetration depth and $C_s = 2680\ m/s$ is the speed of sound in silver. Figure 11 which shows that QM estimated $\tau_{e-ph,QM}$ is much greater than $\tau_c$. The estimation of $\tau_{e-ph,QM}$ comes from the ratio of $C_e$ to $G_{e-ph}$ from the QM calculated result. Therefore, the ablation observed in this paper is not induced by mechanical confinement. The thermal confinement says that the energy transporting from hot electron to cold lattice leads fast thermal expansion of lattice, which agrees with simulation result seen in Fig. 9. The condition for thermal confinement is defined when the timescale of laser pulse duration $t_p$ is smaller than $\tau_{th} = L_p^2/\alpha_{th}$ [51], where $\alpha_{th}$ is the thermal diffusivity of the laser heated system (either lattice subsystem or electron subsystem). For the lattice subsystem, $\tau_{th,l} = 26.5090\ ps$, since the lattice thermal conductivity and heat capacity are treated as constants. For the electron subsystem, $\tau_{th,e}$ is a variable seen in Fig. 11, which is thousands of times longer than $\tau_{th,l}$. In the present paper, $t_p$ is $500\ fs$, which is significantly shorter than the time needing for dissipating the absorbed laser energy by both electron and lattice thermal conduction. Therefore, thermal confinement is the reason leads to ablation of atomic aggregations from the laser irradiated front surface of the silver film.

Table 1. Depth of homogeneous melting $L_{mel}$, number of atomic aggregation $n_{abl}$ and depth of ablation $L_{abl}$.

| $J_{abs}$ (J/cm²) | $L_{mel}$ (nm) | $n_{abl}$ | $L_{abl}$ (nm) |
| --- | --- | --- | --- |
| 0.0750 | N/A | N/A | N/A |
| 0.0813 | 13.7266 | N/A | N/A |
| 0.0875 | 32.6824 | N/A | N/A |
| 0.1000 | 71.9013 | N/A | N/A |
| 0.1125 | 91.5107 | N/A | N/A |
| 0.1188 | 94.7790 | N/A | N/A |
| 0.1250 | 111.1202 | 1 | 65.3648 |
| 0.1500 | 137.2661 | 3 | 91.5107 |
| 0.1750 | 156.8755 | 7 | 104.5837 |
| 0.2000 | 176.4850 | 8 | 88.2425 |

Note: All the values in the table are measured at $t = 160\ ps$. All the depths are measured by taking the initial location of the front surface ($x = 196.0944\ nm$) as the reference point. "N/A" denotes the phenomenon is not observed at given laser fluence.

### 3.4 Depths and thresholds of melting and ablation

Table 1 lists the depth of homogenous melting $L_{mel}$, number of atomic aggregations $n_{abl}$ and depth of ablation $L_{abl}$, which were obtained by performing QM-MD-TTM integrated simulation for $J_{abs}$ ranging from $0.075\ J/cm^2$ to $0.2\ J/cm^2$. $L_{mel}$ is measured as the distance from the initial location of the front film surface ($x = 196.0944\ nm$) to the interface between the homogenous melted region and the heterogeneous melted region. $L_{mel}$ provides quantitative result depth of homogenous melting with the increase of laser fluence in Table 1. At the present, it is still hard to get the depth of melting, since the heterogeneous melting exists in the deeper region below the depth of homogenous melting. $L_{abl}$ is the distance from the initial location of the front surface to location of the remained bulk silver the film after the innermost ablation. As seen in the second column in Table 1, $L_{mel}$ develops deeper for higher $J_{abs}$ of femtosecond laser heating. The estimated ranges for thresholds of melting $J_{mel}$ and ablation $J_{abl}$ are obtained according to the occurring sequence from melting to ablation. It is found that homogenous melting starts when $0.075 \, J/cm^2 < J_{abs} < 0.0813 \, J/cm^2$. Therefore, it can be concluded that $J_{mel}$ locates in this range. The ablation begins when $J_{abl} > 0.1188 \, J/cm^2$. One atomic aggregation is ablated at the $L_{abl} = 65.3648 \, nm$ for $J_{abs} = 0.125 \, J/cm^2$. Hence, the threshold of ablation is in the range $0.1188 \, J/cm^2 < J_{abl} < 0.125 \, J/cm^2$. With the increase of the laser fluence, the number of ablated atomic aggregations increase. Even though the ablation depth for $J_{abs} = 0.2 \, J/cm^2$ is slightly smaller than that for $0.175 \, J/cm^2$, the number of ablated atomic aggregations and melting depth for $J_{abs} = 0.2 \, J/cm^2$ are greater than those for $0.175 \, J/cm^2$, which indicates there is more proportion of laser energy contributing to melting than to ablation for $J_{abs} = 0.2 \, J/cm^2$.

## 4. Conclusions

The laser heating of silver film is simulated under the framework of QM-MD-TTM integrated simulation. The beginning of *ab initio* calculation helps to achieve laser excitation dependent electron thermophysical properties, which ensures the precision of the subsequent MD-TTM coupled simulation. The approach of MD captures the details of atomic motion and provides the detailed information on the laser melting process from homogenous melting to heterogeneous melting. The effective electron-phonon coupling factor at given electron temperature offers dynamic rate of thermal energy transporting from the electron subsystem to lattice subsystem. The fundamental microscopic mechanisms of melting and ablation are investigated and revealed. Right after laser heating, the heterogeneous melting process initiates along tens of nanometers from laser heated surface. Whereas, homogenous melting gradually develops from the front surface to the silver film. A finally stable interface between the homogenous melted region and the heterogeneous melted region is found in the simulation results. The ablation is observed as a result of thermal expansion of the locally superheated silver. Thresholds for the homogenous melting and ablation are determined by means of gradually increasing the absorbed laser fluence.

## Acknowledgement

Support for this work by the U.S. National Science Foundation under grant number CBET-133611 is gratefully acknowledged.